\documentclass[10pt,onecolumn]{IEEEtran}

\usepackage[cmex10]{amsmath}
\usepackage{mathrsfs}
\usepackage{cite}
\usepackage{enumerate}

\usepackage{amssymb,amsthm}
\usepackage{tikz}
\usetikzlibrary{calc}

\newtheorem{lemma}{Lemma}
\newtheorem{thm}{Theorem}

\newtheorem{prop}{Proposition}

\theoremstyle{definition}

\newtheorem{rem}{Remark}

\theoremstyle{remark}

\DeclareMathOperator*{\argmin}{\arg\!\min}
\DeclareMathOperator\E{\mathsf{E}}

\begin{document}
\allowdisplaybreaks

\title{
Minimax Learning for Remote Prediction 
}

\author{\IEEEauthorblockN{Cheuk Ting Li\textsuperscript{*}, Xiugang Wu\textsuperscript{\dag}, Ayfer Ozgur\textsuperscript{\dag}, Abbas El Gamal\textsuperscript{\dag}}\\
	\IEEEauthorblockA{\textsuperscript{*}Department of Electrical Engineering and Computer Sciences\\
		University of California, Berkeley\\
		Email: ctli@berkeley.edu\\
		\textsuperscript{\dag}Department of Electrical Engineering\\
		Stanford University\\
		Email: \{x23wu, aozgur\}@stanford.edu; abbas@ee.stanford.edu}
		\thanks{This paper was presented in part at the IEEE International
		Symposium on Information Theory, Vail, Colorado, USA, June 2018.%
		}
}

\maketitle

\begin{abstract}
The classical problem of supervised learning is to infer an accurate predictor of a target variable $Y$ from a measured variable $X$ by using a finite number of labeled training samples. Motivated by the increasingly distributed nature of data and decision making, in this paper we consider a variation of this classical problem in which the prediction is performed remotely based on a rate-constrained description $M$ of $X$. Upon receiving $M$, the remote node computes an estimate $\hat Y$ of $Y$. We follow the recent minimax approach to study this learning problem and show that it corresponds to a one-shot minimax noisy source coding problem. We then establish information theoretic bounds on the risk-rate Lagrangian cost and a general method to design a near-optimal descriptor-estimator pair, which can be viewed as a rate-constrained analog to the maximum conditional entropy principle used in the classical minimax learning problem. Our results show that a naive estimate-compress scheme for rate-constrained prediction is not in general optimal. 


\end{abstract}


\IEEEpeerreviewmaketitle

\section{Introduction}
\label{sec:introduction}

The classical problem of supervised learning is to infer an accurate predictor of a target variable $Y$ from a measured variable $X$ on the basis of $n$ labeled training samples $\{(X_i, Y_i)\}_{i=1}^n$ independently drawn from an unknown joint distribution $P$. The standard approach for solving this problem in statistical learning theory is empirical risk minimization (ERM). For a given set of allowable predictors and a loss function that quantifies the risk of each predictor, ERM chooses the predictor with minimal risk under the empirical distribution of samples. To avoid overfitting, the set of allowable predictors is restricted to a class with limited complexity. 

Recently, an alternative viewpoint has emerged which seeks distributionally robust predictors. Given the labeled training samples, this approach learns a predictor by minimizing its worst-case risk over an ambiguity distribution set centered at the empirical distribution of samples. In other words, instead of restricting the set of allowable predictors, it aims to avoid overfitting by requiring that the learned predictor performs well under any distribution in a chosen neighborhood of the empirical distribution. This minimax approach has been investigated under different assumptions on how the ambiguity set is constructed, e.g., by restricting the moments \cite{farnia2016minimax}, forming the $f$-divergence balls \cite{
namkoong2017variance} and Wasserstein balls \cite{
lee2017minimax} (see also references therein).


In these previous works, the learning algorithm finds a predictor that acts directly on a fresh (unlabeled) sample $X$ to predict the corresponding target variable $Y$. Often, however the fresh sample $X$ may be only remotely available, and when designing the predictor it is desirable to also take into account  the cost of communicating $X$.  This is motivated by the fact that bandwidth and energy limitations on communication in networks and within multiprocessor systems often impose significant bottlenecks on the performance of algorithms. There are also an increasing number of applications in which data is generated in a distributed manner and it (or features of it) are communicated over bandwidth-limited links to a central processor to perform inference. For instance, applications such as Google Goggles and Siri process the locally collected data on clouds. It is thus important to study prediction in distributed and rate-constrained settings. 

In this paper, we study an extension of the classical learning problem in which given a finite set of training samples, the learning algorithm needs to infer a descriptor-estimator pair with a desired communication rate in between them. This is especially relevant when both $X$ and $Y$ come from a large alphabet or are continuous random variables as in regression problems, so neither the sample $X$ nor its predicted value of $Y$ can be simply communicated in a lossless fashion. We adopt the minimax framework for learning the descriptor-estimator pair. Given a set of labeled training samples, our goal is to find a descriptor-estimator pair by minimizing their resultant worst-case risk over an ambiguity distribution set, where the risk now incorporates both the statistical risk and the communication cost. One of the important conclusions that emerge from the minimax  approach to supervised learning in \cite{farnia2016minimax} is that the problem of finding the predictor with minimal worst-case risk over an ambiguity set can be broken into two smaller steps: (1) find the worst-case distribution in the ambiguity set that maximizes the (generalized) conditional entropy of $Y$ given $X$, and (2) find the optimal predictor under this worst-case distribution. In this paper, we show that an analogous principle approximately holds for rate-constrained prediction. The descriptor-estimator pair with minimal worst-case risk can be found in two steps: (1) find the worst-case distribution in the ambiguity set that maximizes the risk-information Lagrangian cost, and (2) find the optimal descriptor-estimator pair under this worst-case distribution. We then apply our results to characterize the optimal descriptor-estimator pairs for two applications: rate-constrained linear regression and rate-constrained classification. While a simple scheme whereby we first find the optimal predictor ignoring the rate constraint, then compress and communicate the predictor output, is optimal for the linear regression application, we show via the classification application that such an estimate-compress approach is not optimal in general. We show that when prediction is rate-constrained, the optimal descriptor aims to send sufficiently (but not necessarily maximally) informative features of the observed variable, which are at the same time easy to communicate. When applied to the case in which the ambiguity distribution set contains only a single distribution (for example, the true or empirical distribution of $X,Y$) and the loss function for the prediction is logarithmic loss, our results provide a new one-shot operational interpretation of the information bottleneck problem. A key technical ingredient in our results is the strong functional representation lemma (SFRL) developed in \cite{sfrl_isit}, which we use to design the optimal descriptor-estimator pair for the worst-case distribution.

\subsection*{Notation}

We assume that $\log$ is base 2 and the entropy
$H$ is in bits. The length of a variable-length description $M \in \{0,1\}^*$ is denoted as $|M|$.  For random variables $U,V$, denote the joint distribution by $P_{U,V}$ and the conditional distribution of $U$ given $V$ by $P_{U|V}$. For brevity we denote the distribution of $(X,Y)$ as $P$. We write $I_{P}(X;\hat{Y})$ for $I(X;\hat{Y})$ when $(X,Y) \sim P$, and $P_{\hat Y|X}$ is clear from the context.

\section{Problem Formulation}
\label{sec:problemform}
We begin by reviewing the minimax approach to the classical learning problem \cite{farnia2016minimax}.

\subsection{Minimax Approach to Supervised Learning}

Let $X\in \mathcal X$ and $Y\in \mathcal Y$ be jointly distributed random
variables. The problem of statistical learning is to design an
accurate predictor of a target variable $Y$ from  a measured
variable $X$ on the basis of a number of independent training
samples $\{(X_i,Y_i)\}_{i=1}^n$ drawn from an unknown joint distribution. The  standard approach for solving this problem is to use empirical risk minimization (ERM) in which one defines an admissible class of predictors $\mathcal F$ that consists of functions  $f: \mathcal X \to \hat{\mathcal Y}$ (where the reconstruction alphabet $\hat{\mathcal Y}$ can be in general different from  ${\mathcal Y}$) and a loss function $\ell : \hat{\mathcal Y} \times \mathcal Y  \to \mathbb R $.  The risk associated with a predictor $f$ when the underlying joint distribution of $X$ and $Y$ is $P$ is  
\[ 
 L(f,P)\triangleq \E_{P}[\ell(f(X),Y)].
 \] 
ERM simply chooses the predictor $f_n \in\mathcal F$ with minimal risk under the empirical distribution $P_n$ of the training samples. 
 
\begin{figure}[h!]
\centering
\includegraphics[width=0.5\textwidth]{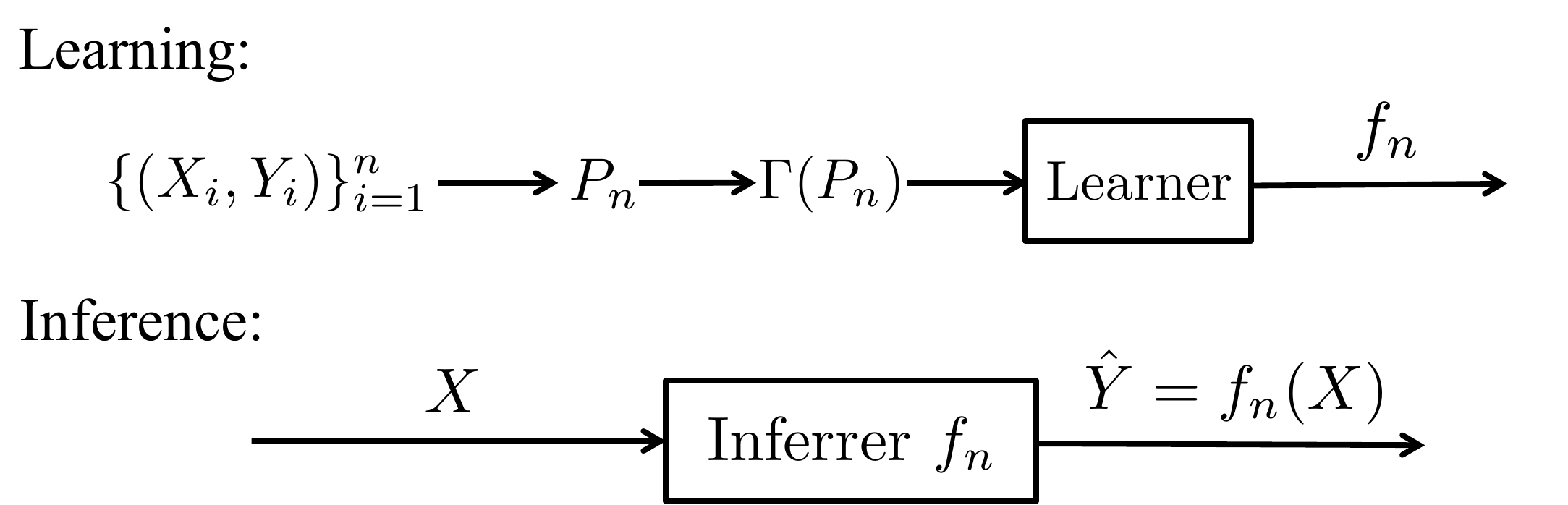}
\caption{Minimax approach to supervised learning.}
\label{F:classical}
\end{figure}

Recently, an alternative approach has emerged which seeks distributionally robust predictors. This approach learns a predictor by  minimizing its worst-case risk over an ambiguity distribution set $\Gamma(P_n)$, i.e.,
\begin{align}
f_n = \argmin_{f}\max_{P\in\Gamma(P_n)} L(f,P),
\end{align} 
where $f$ can be any function and $\Gamma(P_n)$ can be constructed in various ways, e.g., by restricting the moments, forming the $f$-divergence balls or Wasserstein balls.  While in ERM it is important to restrict the set $\mathcal F$ of admissible predictors to a low-complexity class to prevent overfitting, in the minimax approach overfitting is prevented by explicitly requiring that the chosen predictor is distributionally robust. The learned function $f_n$ can be then used for predicting $Y$ when presented with  fresh samples of $X$. The learning and inference phases are illustrated in Figure \ref{F:classical}.

 \subsection{Minimax Learning for Remote Prediction}

\begin{figure}[h!]
\centering
\includegraphics[width=0.5\textwidth]{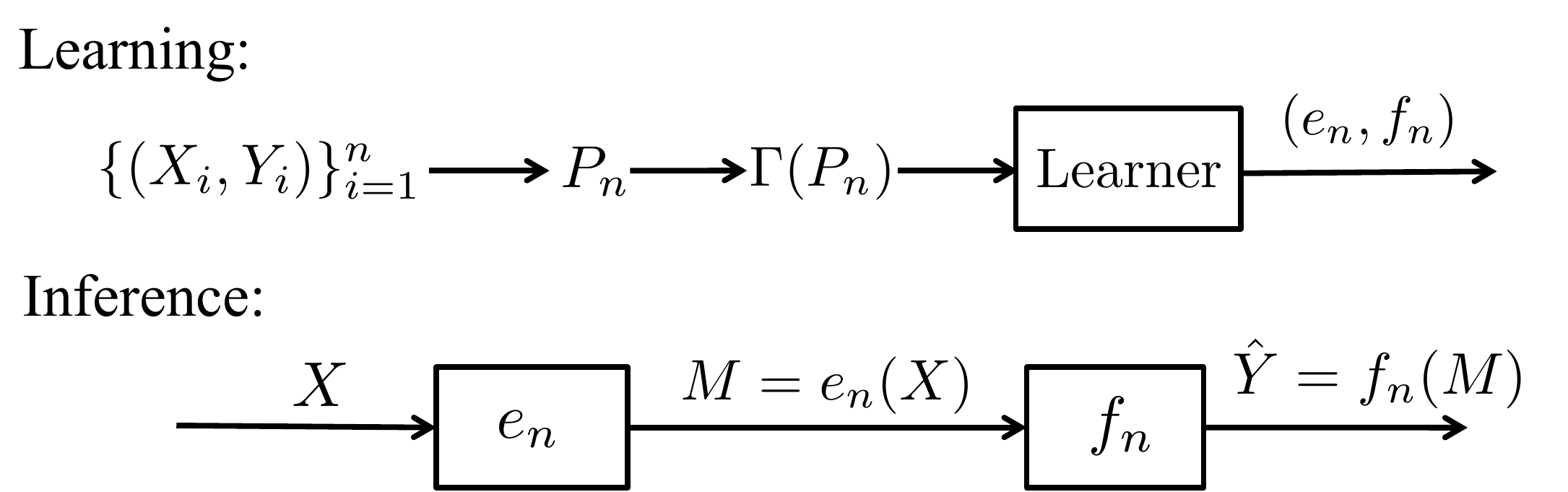}
\caption{Minimax learning for remote prediction.}
\label{F:comm_constraint}
\end{figure}

In this paper, we extend the minimax learning approach to the setting in which the prediction needs to be performed based on a rate-constrained description of $X$. In particular, given a set of finite training samples $\{(X_i,Y_i)\}_{i=1}^n$ independently drawn from an unknown joint distribution $P$, our goal is to learn a pair of functions $(e,f)$, where $e$ is a descriptor used to compress $X$ into $M=e(X)\in \{0,1\}^*$ (a prefix-free code), and   $f$ is an estimator that takes the compression $M$ and generates an estimate $\hat Y$ of $Y$. See Figure~\ref{F:comm_constraint}. 

Let $R(e, P)\triangleq \E_{P}[|e(X)|]$ be the rate of the descriptor $e$ and $L(e,f,P)\triangleq \E_{P}[\ell(f(e(X)),Y)]$ be the risk associated with the descriptor-estimator pair $(e,f)$, when the underlying distribution of  $(X,Y)$ is $P$, and define the risk-rate Lagrangian cost (parametrized by $\lambda>0$) as 
\begin{align}
L_\lambda(e,f,P) = L(e,f,P) + \lambda R(e, P). \label{eqn:lcr}
\end{align}
Note that this cost function takes into account both the resultant statistical prediction risk of $(e,f)$, as well as the communication rate they require. The task of a minimax learner is to find an $(e_n,f_n)$ pair that minimizes the worst-case $L_\lambda(e,f,P)$ over the ambiguity distribution set $\Gamma(P_n)$, i.e.,
\begin{align}
(e_n, f_n) = \argmin_{(e,f)}\max_{P\in\Gamma(P_n)} L_\lambda(e,f,P),
\end{align} 
for an appropriately chosen $\Gamma(P_n)$ centered at the empirical distribution of samples $P_n$. Note that we allow here all possible $(e,f)$ pairs. We also assume that the descriptor and the estimator can use unlimited common randomness $W$ which is independent of the data, i.e.,  $e$ and $f$ can be expressed as functions of $(X,W)$ and $(M,W)$, respectively, and the prefix-free codebook for $M$ can depend on $W$.  The availability of such common randomness can be justified by the fact that in practice, although the inference scheme is one-shot, it is used many times (by the same user and by different users), hence the descriptor and the estimator can share a common randomness seed before communication commences without impacting the communication rate.

\section{Main Results}

We first consider the case where $\Gamma$ consists of a single distribution $P$, which may be the empirical distribution $P_n$ as in ERM. Define the minimax risk-rate cost as
\begin{align}
L^*_\lambda(\Gamma) = \inf_{(e,f)} \sup_{P \in \Gamma} L_\lambda(e, f, P). \label{eqn:mlcr}
\end{align}
While it is difficult to minimize the risk-rate cost~\eqref{eqn:lcr} directly, the minimax risk-rate cost can be bounded in terms of the mutual information between $X$ and $\hat Y$. 
\begin{thm}
\label{thm:fixedp}Let $\Gamma=\{P\}$. Then
\begin{align*}
L_{\lambda}^{*}\ge & \inf_{P_{\hat{Y}|X}}\left(\E\left[\ell(\hat{Y},Y)\right]+\lambda I(X;\hat{Y})\right),\\
L_{\lambda}^{*}\le & \inf_{P_{\hat{Y}|X}}\Big(\E\left[\ell(\hat{Y},Y)\right]   +  \lambda\left( I(X;\hat{Y}) + \log(I(X;\hat{Y}) + 1) + 5\right) \Big) .
\end{align*}
\end{thm}

As in other one-shot compression results (e.g., zero-error compression), there is a gap between the upper and lower bound. While the logarithmic gap in Theorem~\ref{thm:fixedp} is not as small as the 1-bit gap in the zero-error compression, it is dominated by the linear term $I(X;\hat{Y})$ when it is large.

To prove Theorem~\ref{thm:fixedp}, we use the strong functional representation lemma given in~\cite{sfrl_isit} (also see~\cite{harsha2010communication,braverman2014public}): for any random variables $X,\hat{Y}$, there exists random variable $W$ independent of $X$, such that $\hat{Y}$ is a function of $(X,W)$, and
\begin{align}
H(\hat{Y}|W) &\le I(X;\hat{Y}) + \log(I(X;\hat{Y})+1)+4. \label{eqn:sfrl}
\end{align}
Here, $W$ can be intuitively viewed as the part of $\hat{Y}$ which is not contained in $X$. Note that for any $W$  such that $\hat{Y}$ is a function of $(X,W)$ and $W$ is independent of $X$, $H(\hat{Y}|W)\geq I(X;\hat{Y})$.  The statement \eqref{eqn:sfrl} ensures the existence of an $W$, independent of $X$, which comes close to this lower bound, and in this sense it is most informative about $\hat{Y}$. This is critical for the proof of Theorem~\ref{thm:fixedp} as we will see next. Identifying the part of $\hat{Y}$ which is not contained in $X$ allows us to generate and share this part between the descriptor and the estimator ahead of time, eliminating the need to communicate it during the course of inference. To find $W$, we use the Poisson functional representation construction detailed in~\cite{sfrl_isit}.
\medskip

\begin{IEEEproof}[Proof of Theorem~\ref{thm:fixedp}]
Recall that $\hat{Y}=f(e(X,W),W)$. The lower bound follows from the fact that $I_P(X; \hat{Y}) \le H_P(M) \le \E [|M|]$. To establish the upper bound, fix any $P_{\hat{Y}|X}$. Let $W$ be obtained from~\eqref{eqn:sfrl}. Note that $W$ is independent of $X$ and can be generated from a random seed shared between the descriptor and the estimator ahead of time. For a given $w$, take $m=e(x,w)$ to be the Huffman codeword of $\hat{y}(x,w)$ according to the distribution $P_{\hat{Y} | W}(\cdot | w)$ (recall that $\hat{Y}$ is a function of $(X,W)$), and take $f(m,w)$ to be the decoding function of the Huffman code. The expected codeword length
\[
\E[|M|] \le H(\hat{Y}|W)+1 \le I(X;\hat{Y}) + \log(I(X;\hat{Y})+1)+5.
\]
Taking an infimum over all $P_{\hat{Y}|X}$ completes the proof.
\end{IEEEproof}

\begin{rem}
If we consider the logarithmic loss $\ell(\hat{y},y)= -\log \hat{y}(y)$, where $\hat{y}$ is a distribution over $\mathcal{Y}$, then the lower bound in Theorem \ref{thm:fixedp} reduces to
\[
\inf_{P_{U|X}}\left( H(Y|U) + \lambda I(X ; U)\right)   =  H(Y) + \inf_{P_{U|X}}\left(\lambda I(X ; U)  -  I(Y ; U)\right) ,
\]
which is the information bottleneck function~\cite{tishby2000information}. Therefore the setting of remote prediction provides an approximate one-shot operational interpretation of the information bottleneck (up to a logarithmic gap). In~\cite{harremoes2007information, courtade2014multiterminal} it was shown that the asymptotic noisy source coding problem also provides an operational interpretation of the information bottleneck. Our operational interpretation, however, is more satisfying since the feature extraction problem originally considered in \cite{tishby2000information} is by nature one-shot. 
\end{rem}

We now extend Theorem~\ref{thm:fixedp} to the minimax
setting. 
\begin{thm}
\label{thm:minimax_bound}Suppose $\Gamma$ is convex.
Then 
\begin{align*}
L_{\lambda}^{*}\ge & \inf_{P_{\hat{Y}|X}}\sup_{P\in\Gamma}\left(\E_{P}\left[\ell(\hat{Y},Y)\right]+\lambda I_{P}(X;\hat{Y})\right)\\
L_{\lambda}^{*}\le & \inf_{P_{\hat{Y}|X}}\sup_{P\in\Gamma}\Big(\E_{P}\left[\ell(\hat{Y},Y)\right]\\
 & \;\;\;\;+\lambda\left(I_{P}(X;\hat{Y})+2\log(I_{P}(X;\hat{Y})+1)+6\right)\Big).
\end{align*}
\end{thm}
This result is related to minimax noisy source coding~\cite{dembo2003minimax}. The main difference is that we consider the one-shot expected length instead of the asymptotic rate.

To prove this theorem, we first invoke a minimax result for relative
entropy in~\cite{haussler1997minimax} (which generalizes the redundancy-capacity
theorem~\cite{gallager1979source}). Then we apply the following
refined version of the strong functional representation lemma that
is proved in the proof of Theorem 1 in~\cite{sfrl_isit} (also see~\cite{harsha2010communication}).

\begin{lemma}\label{lem:sfrl_div} For any $P_{\hat{Y}|X}$ and $\tilde{P}_{\hat{Y}}$,
there exists random variable $W$, and functions $k(x,w)\in\{1,2,\ldots\}$
and $\hat{y}(k,w)$ such that $\hat{y}(k(x,W),W)\sim P_{\hat{Y}|X}(\cdot|x)$,
and 
\begin{equation}
\E\left[\log k(x,W)\right]\le D\bigl(P_{\hat{Y}|X}(\cdot|x)\,\bigl\Vert\,\tilde{P}_{\hat{Y}}\bigr)+1.6.\label{eq:sfrl_div}
\end{equation}
\end{lemma}

We are now ready to prove Theorem~\ref{thm:minimax_bound}. 
\begin{IEEEproof}
The lower bound follows from $\E_{P}[|M|]\ge H_{P}(M)\ge I_{P}(X;\hat{Y})$.
To prove the upper bound, we fix any $P_{\hat{Y}|X}$, and show
that the following risk-rate cost is achievable:

\begin{align*}
L' & =\sup_{P\in\Gamma}\Big(\E_{P}\left[\ell(\hat{Y},Y)\right]\\
 & \;\;\;+\lambda\left(I_{P}(X;\hat{Y})+2\log(I_{P}(X;\hat{Y})+1)+6\right)\Big).
\end{align*}

\noindent Let
\begin{align*}
g(P,\tilde{P}_{\hat{Y}}) & =\E_{P}\left[\ell(\hat{Y},Y)\right]+\lambda\biggl(\int D\bigl(P_{\hat{Y}|X=x}\,\bigl\Vert\,\tilde{P}_{\hat{Y}}\bigr)dP(x)\\
 & \;\;\;\;\;+2\log\Bigl(\int D\bigl(P_{\hat{Y}|X=x}\,\bigl\Vert\,\tilde{P}_{\hat{Y}}\bigr)dP(x)+1\Bigr)+6\biggr).
\end{align*}

\noindent Note that $g$ is concave in $P$ for fixed $\tilde{P}_{\hat{Y}}$ since $\E_{P}\left[\ell(\hat{Y},Y)\right]$
and $\int D\bigl(P_{\hat{Y}|X=x}\,\bigl\Vert\,\tilde{P}_{\hat{Y}}\bigr)dP(x)$
are linear in $P$. Also $g$ is quasiconvex in $\tilde{P}_{\hat{Y}}$ for fixed $P$
since $\int D\bigl(P_{\hat{Y}|X=x}\,\bigl\Vert\,\tilde{P}_{\hat{Y}}\bigr)dP(x)$
is convex in $\tilde{P}_{\hat{Y}}$, and is lower semicontinuous in
$\tilde{P}_{\hat{Y}}$ since $D\bigl(P_{\hat{Y}|X=x}\,\bigl\Vert\,\tilde{P}_{\hat{Y}}\bigr)$
is lower semicontinuous with respect to the topology of weak convergence~\cite{posner1975minimum},
and hence $\int D\bigl(P_{\hat{Y}|X=x}\,\bigl\Vert\,\tilde{P}_{\hat{Y}}\bigr)dP(x)$
is lower semicontinuous by Fatou's lemma.

Write $P_{\hat{Y}|X} \circ P$ for the distribution of $\hat{Y}$ when $(X,Y)\sim P$
and $\hat{Y}|\{X=x\}\sim P_{\hat{Y}|X}(\cdot|x)$. Let $\Gamma_{\hat{Y}}=\{P_{\hat{Y}|X} \circ P:\,P\in\Gamma\}$ and $\overline{\Gamma_{\hat{Y}}}$
be the closure of $\Gamma_{\hat{Y}}$ in the topology of weak convergence.
It can be shown using the same arguments as in~\cite{haussler1997minimax}
(on $g$ instead of relative entropy, and using Sion's minimax theorem~\cite{sion1958general}
instead of Lemma 2 in~\cite{haussler1997minimax}) that if $\Gamma_{\hat{Y}}$
is uniformly tight, then there exists $P_{\hat{Y}}^{*}\in\overline{\Gamma_{\hat{Y}}}$
such that
\[
\sup_{P\in\Gamma}g(P,\tilde{P}_{\hat{Y}}^{*})=\sup_{P\in\Gamma}\inf_{\tilde{P}_{\hat{Y}}}g(P,\tilde{P}_{\hat{Y}})=L'.
\]
If $\Gamma_{\hat{Y}}$ is not uniformly tight, then by Lemma 4 in~\cite{haussler1997minimax}, $\sup_{P\in\Gamma}\inf_{\tilde{P}_{\hat{Y}}}\int D\bigl(P_{\hat{Y}|X=x}\,\bigl\Vert\,\tilde{P}_{\hat{Y}}\bigr)dP(x)=\infty$,
and hence $L'=\sup_{P\in\Gamma}\inf_{\tilde{P}_{\hat{Y}}}g(P,\tilde{P}_{\hat{Y}})=\infty$.

Applying Lemma~\ref{lem:sfrl_div} to $P_{\hat{Y}|X}$, $P_{\hat{Y}}^{*}$
we obtain $W$ independent of $X$, random variable $K=k(X,W)\in\{1,2,\ldots\}$,
and $\hat{Y}=\hat{y}(K,W)$ following the conditional distribution
$P_{\hat{Y}|X}$, and 
\[
\E\left[\log K\,|\,X=x\right]\le D\bigl(P_{\hat{Y}|X}\,\bigl\Vert\,P_{\hat{Y}}^{*}\,\bigr|\,X=x\bigr)+1.6
\]
for any $x$. Then we use Elias delta code~\cite{elias1975} for
$K$ to produce $M$. Note that the average length of the Elias delta
code is upper bounded by $\log K+2\log\left(\log K+1\right)+1$. Hence,
we have 
\begin{align*}
\E_{P}\left[|M|\right] & \le \E_{P}\left[\log K\right]+2\log\left(\E_{P}\left[\log K\right]+1\right)+1\\
 & \le\int D\bigl(P_{\hat{Y}|X=x}\,\bigl\Vert\,P_{\hat{Y}}^{*}\bigr)dP(x)\\
 & \;\;+2\log\left(\int D\bigl(P_{\hat{Y}|X=x}\,\bigl\Vert\,P_{\hat{Y}}^{*}\bigr)dP(x)+1\right)+6.
\end{align*}
Hence 
\begin{align*}
\tilde{L}_{\lambda}^{*}\le & \sup_{P\in\Gamma}\left(\E_{P}\left[\ell(\hat{Y},Y)+\lambda|M|\right]\right)\le\sup_{P\in\Gamma}g(P,P_{\hat{Y}}^{*})\le L'.
\end{align*}
\end{IEEEproof}


Theorem \ref{thm:minimax_bound} suggest that we can simplify the analysis of the 
risk-rate cost~\eqref{eqn:lcr} $L_{\lambda}=\E_{P}\left[\ell(\hat{Y},Y)\right]+\lambda\E_{P}\left[|M|\right]$ by replacing the rate $\E_{P}\left[|M|\right]$ with the mutual information $I_{P}(X;\hat{Y})$. Define the \emph{risk-information cost} as
\begin{align}
\tilde{L}_{\lambda}(P_{\hat{Y}|X},P)=\E_{P}\big[\ell(\hat{Y},Y)\big]+\lambda I_{P}(X;\hat{Y}). \label{eqn:lir}
\end{align}
Theorem~\ref{thm:minimax_bound} implies that the minimax risk-rate cost $L^*_\lambda$ can be approximated by the \emph{minimax risk-information cost}
\begin{align}
\tilde{L}^*_\lambda(\Gamma) = \inf_{P_{\hat{Y}|X}} \sup_{P \in \Gamma} \tilde{L}_\lambda(P_{\hat{Y}|X}, P), \label{eqn:mlir}
\end{align}
within a logarithmic gap. Theorem~\ref{thm:minimax_bound} can also be stated in the following slightly weaker form
\[
\tilde{L}^*_\lambda \le L^*_\lambda \le \tilde{L}^*_\lambda + 2 \lambda \log(\lambda^{-1}\tilde{L}^*_\lambda +1) + 7 \lambda.
\]
The risk-information cost has more desirable properties than the risk-rate cost.
For example, it is convex in $P_{\hat{Y}|X}$ for fixed $P$,
and concave in $P$ for fixed $P_{\hat{Y}|X}$. This allows us to
exchange the infimum and supremum in Theorem \ref{thm:minimax_bound} by Sion's minimax theorem~\cite{sion1958general},
which gives the following proposition.

\medskip{}

\begin{prop}
\label{prop:minimax_exchange}Suppose $\mathcal{X}$, $\mathcal{Y}$ and $\hat{\mathcal{Y}}$
are finite, $\Gamma$ is convex and closed, and $\lambda\ge0$, then
\[
\tilde{L}^*_\lambda(\Gamma) =\inf_{P_{\hat{Y}|X}}\sup_{P\in\Gamma}\tilde{L}_{\lambda}(P_{\hat{Y}|X},P)=\sup_{P\in\Gamma}\inf_{P_{\hat{Y}|X}}\tilde{L}_{\lambda}(P_{\hat{Y}|X},P).
\]
Moreover, there exists $P_{\hat{Y}|X}^{*}$
attaining the infimum in the left hand side, which also attains the
infimum on the right hand side when $P$ is fixed to $P^{*}$, the distribution
that attains the supremum on the right hand side.
\end{prop}

\medskip{}

Proposition \ref{prop:minimax_exchange} means that in order to design a robust descriptor-estimator pair that work for any $P\in \Gamma$, we only need to design them according to the worst-case distribution $P^*$ as follows.

\smallskip
\noindent \textbf{Principle of maximum risk-information cost}: Given
a convex and closed $\Gamma$, we design the descriptor-estimator pair based on the worst-case distribution
\[
P^{*}=\underset{P\in\Gamma}{\arg\max}\inf_{P_{\hat{Y}|X}}\tilde{L}_{\lambda}(P_{\hat{Y}|X},P).
\]
We then find $P_{\hat{Y}|X}$ that minimizes $\tilde{L}_{\lambda}(P_{\hat{Y}|X},P^{*})$ and design the descriptor-estimator pair accordingly, e.g. using Lemma~\ref{lem:sfrl_div} on $P_{\hat{Y}|X}$ and the induced distribution $P_{\hat{Y}}^{*}$ from $P_{\hat{Y}|X}$ and $P^{*}$.


\section{Applications}

\subsection{Rate-constrained Minimax Linear Regression}
Suppose $\mathbf{X} \in \mathbb{R}^d$, $Y \in \mathbb{R}$, $\ell(\hat{y},y)=(y-\hat{y})^2$ is the mean-squared loss, and we observe the data $\{(\mathbf{X}_i,Y_i)\}_{i=1}^n$. Take $\Gamma$ to be the set of distributions with the same first and second moments as given by the empirical distribution, i.e.,
\begin{align}
\Gamma & =\big\{ P_{\mathbf{X}Y}:\,\E[\mathbf{X}]=\boldsymbol{\mu}_\mathbf{X},\,\E[Y]=\mu_Y,\,\mathrm{Var}[\mathbf{X}]=\Sigma_{\mathbf{X}}, \nonumber \\
& \;\;\;\;\;\;\;\;\;\;\;\;\; \mathrm{Var}[Y]=\sigma_{Y}^{2},\,\mathrm{Cov}[\mathbf X,Y]=C_{\mathbf{X}Y}\big\}, \label{eqn:lmmse_set}
\end{align}
where $\boldsymbol{\mu}_\mathbf{X},\mu_Y,\Sigma_{\mathbf{X}},\sigma_{Y}^{2},C_{\mathbf{X}Y}$ are the corresponding statistics of the empirical distribution. The following proposition shows that $P^{*}$ is Gaussian.
\begin{prop}[Linear regression with rate constraint]\label{prop:linreg} Consider mean-squared
	loss and define $\Gamma$ as in~\eqref{eqn:lmmse_set}. Then the minimax risk-information cost~\eqref{eqn:mlir} is
	\begin{align}
	\tilde{L}^*_\lambda &=  \begin{cases}
	\sigma_{Y}^{2} -C_{\mathbf{X}Y}^{T}\Sigma_{\mathbf{X}}^{-1}C_{\mathbf{X}Y} +\frac{\lambda}{2}\log\frac{2 e C_{\mathbf{X}Y}^{T}\Sigma_{\mathbf{X}}^{-1}C_{\mathbf{X}Y}}{\lambda \log e} & \text{if}\; \frac{\lambda \log e}{2}<C_{\mathbf{X}Y}^{T}\Sigma_{\mathbf{X}}^{-1}C_{\mathbf{X}Y}\\
	\sigma_{Y}^{2} & \text{if}\;\frac{\lambda \log e}{2} \ge C_{\mathbf{X}Y}^{T}\Sigma_{\mathbf{X}}^{-1}C_{\mathbf{X}Y},
	\end{cases} \label{E:minimax_cost}
	\end{align}
	where the optimal $P^*_{\mathbf X Y}$ is Gaussian with its mean and covariance matrix specified in \eqref{eqn:lmmse_set}, and the optimal estimate 
	\[
	\hat{Y}=\begin{cases}
	a C_{\mathbf{X}Y}^{T}\Sigma_{\mathbf{X}}^{-1}\mathbf{X}+b+Z & \text{if}\;\frac{\lambda \log e}{2}<C_{\mathbf{X}Y}^{T}\Sigma_{\mathbf{X}}^{-1}C_{\mathbf{X}Y}\\
	\mu_Y & \text{if}\;\frac{\lambda \log e}{2} \ge C_{\mathbf{X}Y}^{T}\Sigma_{\mathbf{X}}^{-1}C_{\mathbf{X}Y},
	\end{cases}
	\]
	where
	\[
	a = 1-\frac{ \lambda \log e}{2 C_{\mathbf{X}Y}^{T}\Sigma_{\mathbf{X}}^{-1}C_{\mathbf{X}Y}},\;
	b = \mu_Y - a C_{\mathbf{X}Y}^{T}\Sigma_{\mathbf{X}}^{-1}\boldsymbol{\mu}_\mathbf{X},
	\]
	and $Z\sim N(0,\,\sigma_{Z}^{2})$ is independent of $\mathbf X$ with $\sigma_{Z}^{2}=\frac{\lambda a\log e}{2}$.  
\end{prop}
 
Note that this setting does not satisfy the conditions in Proposition~\ref{prop:minimax_exchange}. We directly analyze \eqref{eqn:mlir} to obtain the optimal $P^*_{\mathbf X Y}$. Given the optimal $P^*_{\mathbf X Y}$,   Theorem~\ref{thm:minimax_bound} and Lemma~\ref{lem:sfrl_div} can be used to construct the scheme.
Operationally, $e_n(x,w)$ is a random quantizer of $a C_{\mathbf{X}Y}^{T}\Sigma_{\mathbf{X}}^{-1}\mathbf{X}+b$ such that the quantization noise follows $\mathrm{N}(0,\sigma_{Z}^{2})$. With this natural choice of the ambiguity set, our formulation recovers a compressed version of the familiar MMSE estimator.

\begin{figure}[h!]
\centering
\includegraphics[width=0.55\textwidth]{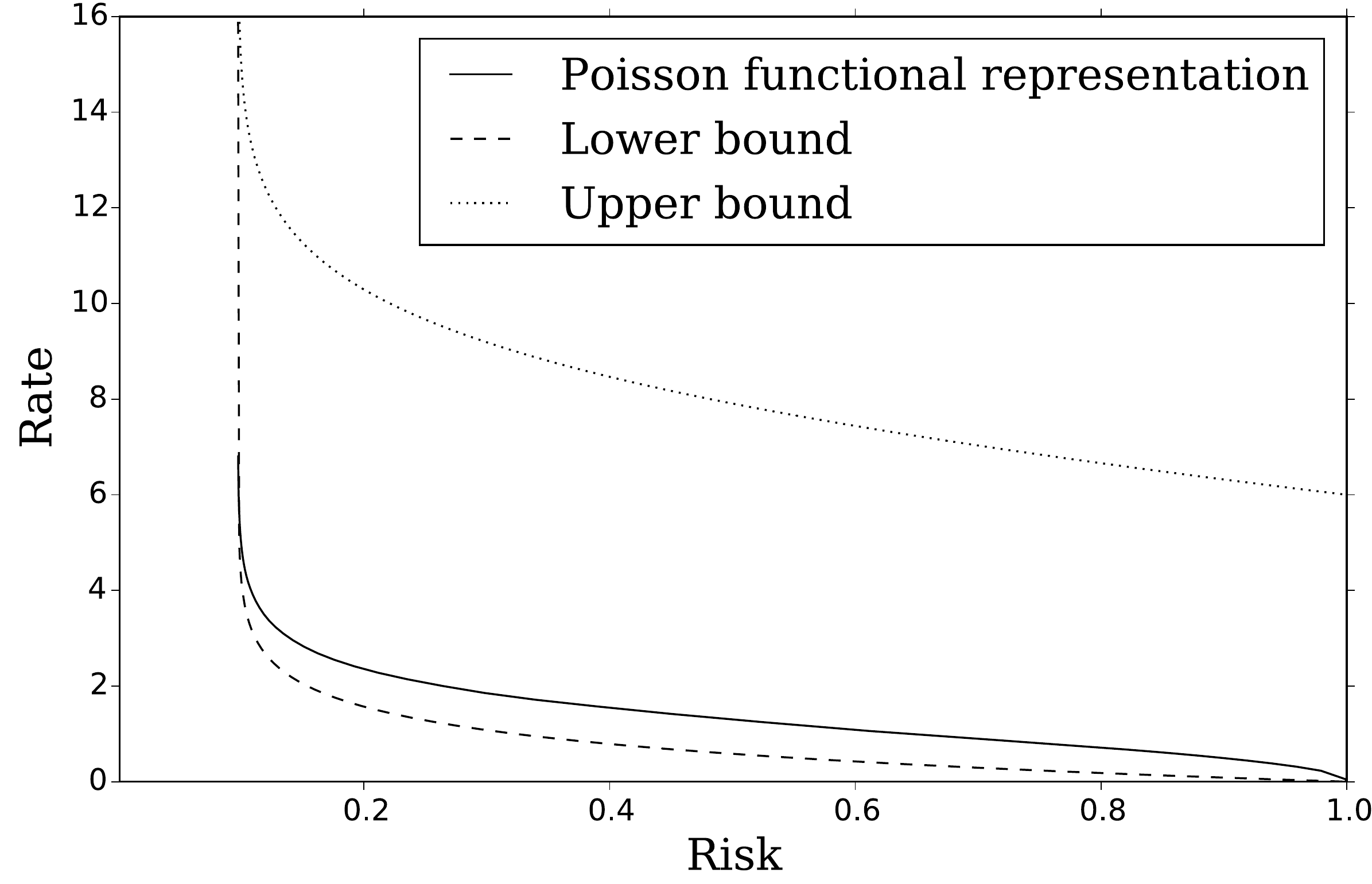}
\caption{Tradeoff between the rate and the risk in rate-constrained minimax linear regression.}
\label{fig:linreg}
\end{figure}

Figure \ref{fig:linreg} plots the tradeoff between the rate and the risk when $d = 1$, $\mu_X=\mu_Y=0$, $\sigma_X^2=\sigma_Y^2 =1$, $C_{XY} = 0.95$ for the scheme constructed using the Poisson functional representation in~\cite{sfrl_isit}, with the lower bound given by the minimax risk-information cost $\tilde{L}^*_\lambda$, and the upper bound given in Theorem~\ref{thm:minimax_bound}.

\begin{IEEEproof}[Proof of Proposition \ref{prop:linreg}]
	Without loss of generality, assume $\boldsymbol{\mu}_{\mathbf X}= \mathbf 0$ and $\mu_{Y}=0$.
	We first prove  ``$\leq$ ''  in \eqref{E:minimax_cost}. For this, fix $P_{\hat{Y}|\mathbf X}$ as given in the proposition and consider any $P\in\Gamma$.  When $\frac{\lambda \log e}{2} < C_{\mathbf{X}Y}^{T}\Sigma_{\mathbf{X}}^{-1}C_{\mathbf{X}Y}$, we have
	\begin{align*}
	\E_{P}\left[\ell(\hat{Y},Y)\right]   &=\E_{P}\left[(\hat{Y}-Y)^{2}\right]\\
	& \le \sigma_{Y}^{2}+ \frac{\lambda \log e}{2}-C_{\mathbf{X}Y}^{T}\Sigma_{\mathbf{X}}^{-1}C_{\mathbf{X}Y}, \text{ and}\\
	I_{P}(\mathbf{X};\hat{Y}) & = h(\hat Y)-h(\hat Y|\mathbf{X})\\
	& \le \frac{1}{2}\log\left(\frac{ 2 C_{\mathbf{X}Y}^{T}\Sigma_{\mathbf{X}}^{-1}C_{\mathbf{X}Y} }{  \lambda   \log e }\right).
	\end{align*}
	Therefore,
	\begin{align*}
	\inf_{P_{\hat{Y}|X}}\sup_{P\in\Gamma}\left(\E_{P}\left[\ell(\hat{Y},Y)\right]+\lambda I_{P}(X;\hat{Y})\right) & \leq \text{R.H.S. of \eqref{E:minimax_cost}}.   
	\end{align*}
	It can also be checked that the above relation holds when $\frac{\lambda \log e}{2} \ge C_{\mathbf{X}Y}^{T}\Sigma_{\mathbf{X}}^{-1}C_{\mathbf{X}Y}$, and thus we have proved ``$\leq$ ''  in \eqref{E:minimax_cost}.
	
	To prove  ``$\geq$ ''  in \eqref{E:minimax_cost}, fix a Gaussian $P_{\mathbf X Y}$ with its mean and covariance matrix specified in \eqref{eqn:lmmse_set} and consider an arbitrary $P_{\hat Y|\mathbf X}$. We have
	\begin{align*}
	&\E_{P}\left[\ell(\hat{Y},Y)\right]   =\E_{P}\left[(Y-\hat{Y})^{2}\right]\\
	& =\sigma_{Y}^{2}-C_{\mathbf{X}Y}^{T}\Sigma_{\mathbf{X}}^{-1}C_{\mathbf{X}Y}+\E_{P}\left[(\hat{Y}-C_{\mathbf{X}Y}^{T}\Sigma_{\mathbf{X}}^{-1}\mathbf{X})^{2}\right],\text{ and}\\
	& I_{P}(X;\hat{Y})  =I_{P}\left(C_{\mathbf{X}Y}^{T}\Sigma_{\mathbf{X}}^{-1}\mathbf{X}\,;\,\hat{Y}\right)\\
	& \ge h\left(C_{\mathbf{X}Y}^{T}\Sigma_{\mathbf{X}}^{-1}\mathbf{X}\right)-h\left(C_{\mathbf{X}Y}^{T}\Sigma_{\mathbf{X}}^{-1}\mathbf{X}-\hat{Y}\right)\\
	& \ge \frac{1}{2}\log C_{\mathbf{X}Y}^{T}\Sigma_{\mathbf{X}}^{-1}C_{\mathbf{X}Y} -\frac{1}{2}\log\E_{P}\left[(\hat{Y}-C_{\mathbf{X}Y}^{T}\Sigma_{\mathbf{X}}^{-1}\mathbf{X})^{2}\right].
	\end{align*}
	Letting $\gamma=\E_{P}\left[(\hat{Y}-C_{\mathbf{X}Y}^{T}\Sigma_{\mathbf{X}}^{-1}\mathbf{X})^{2}\right]$, we  have
	\begin{align*}
	& \E_{P}\left[\ell(\hat{Y},Y)\right]+\lambda I_{P}(X;\hat{Y})\\
	& \ge\sigma_{Y}^{2}-C_{\mathbf{X}Y}^{T}\Sigma_{\mathbf{X}}^{-1}C_{\mathbf{X}Y}+ \frac{\lambda}{2}\log C_{\mathbf{X}Y}^{T}\Sigma_{\mathbf{X}}^{-1}C_{\mathbf{X}Y} + \gamma -  \frac{\lambda \log\gamma}{2} \\
	& \ge  \text{R.H.S. of \eqref{E:minimax_cost}},
	\end{align*}
	where the second inequality follows by evaluating the minimum value of $\gamma -  \frac{\lambda \log\gamma}{2}$. Combing this with the above completes the proof of Proposition~\ref{prop:linreg}.
\end{IEEEproof}

The optimal scheme in the above example corresponds to compressing and communicating the minimax optimal rate-unconstrained predictor $\bar{Y} = C_{\mathbf{X}Y}^{T}\Sigma_{\mathbf{X}}^{-1}(\mathbf{X} -\boldsymbol{\mu}_\mathbf{X})+\mu_Y$, since the optimal $\hat{Y}$ can be obtained from $\bar{Y}$ by shifting, scaling and adding noise. This estimate-compress approach can be thought as a \emph{separation} scheme, since we first optimally estimate $\bar{Y}$, then optimally communicate it while satisfying the rate constraint. In the next application, we show that such separation is not optimal in general.

\subsection{Rate-constrained Minimax Classification}
We assume $\mathcal{Y}=\hat{\mathcal{Y}}=\{1,\ldots,k\}$
and $\mathcal{X}$ are finite, $\ell(\hat{y},y)=\mathbf{1}\{\hat{y}\neq y\}$,
and $\Gamma$ is closed and convex. The following proposition gives
the minimax risk-information cost and the optimal estimator.
\begin{prop}
	\label{prop:classification} Consider the setting described above.
	The minimax risk-information cost is given by 
	\[
	\tilde{L}_{\lambda}^{*}=\sup_{P\in\Gamma}\biggl(1+\lambda\inf_{\tilde{P}_{\hat{Y}}}\E_{P}\biggl(-\log\sum_{y}2^{\lambda^{-1}P_{Y|X}(y|X)}\tilde{P}_{\hat{Y}}(y)\biggr)\biggr),
	\]
	the worst-case distribution $P^{*}$ is the one attaining the supremum,
	and the optimal estimator is given by $P_{\hat{Y}|X}^{*}(\hat{y}|x)\propto2^{\lambda^{-1}P_{Y|X}^{*}(\hat{y}|x)}\tilde{P}_{\hat{Y}}^{*}(\hat{y})$,
	where $\tilde{P}_{\hat{Y}}^{*}$ attains the infimum (when $P=P^*$), and $P_{Y|X}^{*}$
	is obtained from $P^{*}$. 
	
	In particular, if $\Gamma$ is symmetric for different values of $Y$
	(i.e., for any $y_{1},y_{2}\in\mathcal{Y}$, there exists permutation
	$\pi$ of $\mathcal{Y}$, $\tau$ of $\mathcal{X}$ such that $\pi(y_{1})=y_{2}$
	and $P_{X,Y}\in\Gamma\Leftrightarrow P_{\tau(X),\pi(Y)}\in\Gamma$),
	\[
	\tilde{L}_{\lambda}^{*}=\sup_{P\in\Gamma}\biggl(1+\lambda\log k-\lambda\E_{P}\Bigl(\log\sum_{y}2^{\lambda^{-1}P_{Y|X}(y|X)}\Bigr)\biggr).
	\]
\end{prop}
We can see that when $\lambda\to0$, $P_{\hat{Y}|X}^{*}$ tends to
the maximum a posteriori estimator (under $\bar{P}^{*}$, the worst-case
distribution when $\lambda=0$).
\newpage
\begin{IEEEproof}
	Assume $\Gamma$ is closed and convex. By Proposition \ref{prop:minimax_exchange},
	the minimax rate-information cost is $\tilde{L}_{\lambda}^{*}=\sup_{P\in\Gamma}\inf_{P_{\hat{Y}|X}}\tilde{L}_{\lambda}(P_{\hat{Y}|X},P)$,
	where
	\begin{align*}
	& \inf_{P_{\hat{Y}|X}}\tilde{L}_{\lambda}(P_{\hat{Y}|X},P)\\
	& =\inf_{P_{\hat{Y}|X}}\biggl(\E_{P}\left[\ell(\hat{Y},Y)\right]+\lambda I_{P}(X;\hat{Y})\biggr)\\
	& =\inf_{P_{\hat{Y}|X}}\biggl(P\{\hat{Y}\neq Y\}+\lambda\inf_{\tilde{P}_{\hat{Y}}}\int D\bigl(P_{\hat{Y}|X=x}\,\bigl\Vert\,\tilde{P}_{\hat{Y}}\bigr)dP(x)\biggr)\\
	& =\inf_{\tilde{P}_{\hat{Y}},P_{\hat{Y}|X}}\biggl(P\{\hat{Y}\neq Y\}+\lambda\int D\bigl(P_{\hat{Y}|X=x}\,\bigl\Vert\,\tilde{P}_{\hat{Y}}\bigr)dP(x)\biggr)\\
	& =1+\lambda\inf_{\tilde{P}_{\hat{Y}},P_{\hat{Y}|X}}\E_{P}\left(\sum_{y}P_{\hat{Y}|X}(y|X)\left(\log\frac{P_{\hat{Y}|X}(y|X)}{\tilde{P}_{\hat{Y}}(y)}-\lambda^{-1}P_{Y|X}(y|X)\right)\right)\\
	& =1+\lambda\inf_{\tilde{P}_{\hat{Y}}}\inf_{P_{\hat{Y}|X}}\E_{P}\left(\sum_{y}P_{\hat{Y}|X}(y|X)\left(\log\frac{P_{\hat{Y}|X}(y|X)}{2^{\lambda^{-1}P_{Y|X}(y|X)}\tilde{P}_{\hat{Y}}(y)/\sum_{y'}2^{\lambda^{-1}P_{Y|X}(y'|X)}\tilde{P}_{\hat{Y}}(y')}\right)-\log\sum_{y}2^{\lambda^{-1}P_{Y|X}(y|X)}\tilde{P}_{\hat{Y}}(y)\right)\\
	& \stackrel{(a)}{=}1+\lambda\inf_{\tilde{P}_{\hat{Y}}}\E_{P}\biggl(-\log\sum_{y}2^{\lambda^{-1}P_{Y|X}(y|X)}\tilde{P}_{\hat{Y}}(y)\biggr),
	\end{align*}
	where (a) is due to that relative entropy is nonnegative, and equality
	is attained when $P_{\hat{Y}|X}(y|x)\propto2^{\lambda^{-1}P_{Y|X}(y|X)}\tilde{P}_{\hat{Y}}(y)$. 
	
	Next we consider the case in which $\Gamma$ is symmetric. Consider the
	minimax rate-information cost 
	\[
	\tilde{L}_{\lambda}^{*}=\inf_{P_{\hat{Y}|X}}\sup_{P\in\Gamma}\tilde{L}_{\lambda}(P_{\hat{Y}|X},P)=\inf_{P_{\hat{Y}|X}}\sup_{P\in\Gamma}\biggl(\E_{P}\left[\ell(\hat{Y},Y)\right]+\lambda I_{P}(X;\hat{Y})\biggr).
	\]
	For any $i,j\in\mathcal{Y}=\{1,\ldots,k\}$, let $\pi_{ij}$ be the
	permutation over $\mathcal{Y}$ such that $\pi_{ij}(i)=j$ and let
	$\tau_{ij}$ be the corresponding permutation over $\mathcal{X}$
	in the symmetry assumption. Since the function 
	\[
	P_{\hat{Y}|X}\,\mapsto\,\sup_{P\in\Gamma}\tilde{L}_{\lambda}(P_{\hat{Y}|X},P)
	\]
	is convex and symmetric about $\pi_{ij}$ and $\tau_{ij}$ (i.e.,
	$\sup_{P\in\Gamma}\tilde{L}_{\lambda}(P_{\hat{Y}|X},P)=\sup_{P\in\Gamma}\tilde{L}_{\lambda}(P_{\pi_{ij}\hat{Y}|\tau_{ij}X},P)$),
	to find its infimum, we only need to consider $P_{\hat{Y}|X}$'s satisfying
	$P_{\hat{Y}|X}=P_{\pi_{ij}\hat{Y}|\tau_{ij}X}$ for all $i,j$ (if
	not, we can instead consider the average of $P_{\pi_{ij}^{a}\hat{Y}|\tau_{ij}^{a}X}$
	for $a$ from 1 up to the product of the periods of $\pi_{ij}$ and
	$\tau_{ij}$, which gives a value of the function not larger than
	that of $P_{\hat{Y}|X}$). For brevity we say $P_{\hat{Y}|X}$ is
	symmetric if it satisfies this condition.
	
	Fix any symmetric $P_{\hat{Y}|X}$. Since the function 
	\[
	P\,\mapsto\,\tilde{L}_{\lambda}(P_{\hat{Y}|X},P)
	\]
	is concave and symmetric about $\pi_{ij}$ and $\tau_{ij}$ (i.e.,
	$\tilde{L}_{\lambda}(P_{\hat{Y}|X},P_{X,Y})=\tilde{L}_{\lambda}(P_{\hat{Y}|X},P_{\tau_{ij}X,\pi_{ij}Y})$),
	to find its supremum, we only need to consider symmetric $P$'s. Hence,
	\begin{align*}
	\tilde{L}_{\lambda}^{*} & =\inf_{P_{\hat{Y}|X}\,\mathrm{symm.}}\,\sup_{P\in\Gamma\,\mathrm{symm.}}\tilde{L}_{\lambda}(P_{\hat{Y}|X},P)\\
	& =\inf_{P_{\hat{Y}|X}\,\mathrm{symm.}}\,\sup_{P\in\Gamma\,\mathrm{symm.}}\biggl(\E_{P}\left[\ell(\hat{Y},Y)\right]+\lambda I_{P}(X;\hat{Y})\biggr)\\
	& =\inf_{P_{\hat{Y}|X}\,\mathrm{symm.}}\,\sup_{P\in\Gamma\,\mathrm{symm.}}\biggl(P\{\hat{Y}\neq Y\}+\lambda(\log k-H_{P}(\hat{Y}|X))\biggr)\\
	& =1+\lambda\log k+\lambda\inf_{P_{\hat{Y}|X}\,\mathrm{symm.}}\sup_{P\in\Gamma\,\mathrm{symm.}}\E_{P}\left(\sum_{y}P_{\hat{Y}|X}(y|X)\left(\log P_{\hat{Y}|X}(y|X)-\lambda^{-1}P_{Y|X}(y|X)\right)\right)\\
	& =1+\lambda\log k+\lambda\inf_{P_{\hat{Y}|X}\,\mathrm{symm.}}\sup_{P\in\Gamma\,\mathrm{symm.}}\E_{P}\left(\sum_{y}P_{\hat{Y}|X}(y|X)\log\frac{P_{\hat{Y}|X}(y|X)}{2^{\lambda^{-1}P_{Y|X}(y|X)}/\sum_{y'}2^{\lambda^{-1}P_{Y|X}(y'|X)}}-\log\sum_{y}2^{\lambda^{-1}P_{Y|X}(y|X)}\right)\\
	& \ge1+\lambda\log k+\lambda\inf_{P_{\hat{Y}|X}\,\mathrm{symm.}}\sup_{P\in\Gamma\,\mathrm{symm.}}\E_{P}\left(-\log\sum_{y}2^{\lambda^{-1}P_{Y|X}(y|X)}\right)\\
	& =\sup_{P\in\Gamma\,\mathrm{symm.}}\left(1+\lambda\log k-\lambda\E_{P}\log\sum_{y}2^{\lambda^{-1}P_{Y|X}(y|X)}\right),
	\end{align*}
	where the inequality is because relative entropy is nonnegative (and
	equality is attained when $P_{\hat{Y}|X}(y|x)\propto2^{\lambda^{-1}P_{Y|X}(y|x)}$).
	Note that 
	\[
	1+\lambda\log k-\lambda\E_{P}\log\sum_{y}2^{\lambda^{-1}P_{Y|X}(y|X)}=\inf_{P_{\hat{Y}|X}}\biggl(P\{\hat{Y}\neq Y\}+\lambda(\log k-H_{P}(\hat{Y}|X))\biggr)
	\]
	is an infimum of affine functions of $P$, hence it is concave in
	$P$. Also it is symmetric about $\pi$ and $\tau$, hence 
	\begin{align*}
	\tilde{L}_{\lambda}^{*} & \ge\sup_{P\in\Gamma\,\mathrm{symm.}}\left(1+\lambda\log k-\lambda\E_{P}\log\sum_{y}2^{\lambda^{-1}P_{Y|X}(y|X)}\right)\\
	& =\sup_{P\in\Gamma}\left(1+\lambda\log k-\lambda\E_{P}\log\sum_{y}2^{\lambda^{-1}P_{Y|X}(y|X)}\right).
	\end{align*}
	The other direction follows from setting $\tilde{P}_{\hat{Y}}(y)=1/k$.
\end{IEEEproof}

To show that the estimate-compress approach is not always optimal, 
let $\ell(\hat{y},y)=\mathbf{1}\{\hat{y}\neq y\}$, $\mathcal{Y}=\mathcal{Y}_{1}\cup\mathcal{Y}_{2}$,
where $\mathcal{Y}_{1}\cap\mathcal{Y}_{2}=\emptyset$ and $|\mathcal{Y}_{i}|=k_{i}$
is finite. Let $\Gamma=\{P\}$, where $P$ is such that $(X_{1},X_{2})\sim\mathrm{Unif}(\mathcal{Y}_{1}\times\mathcal{Y}_{2})$,
and $Y=X_{i}$ with probability $q_{i}$ for $i=1,2$. By Proposition~\ref{prop:classification},
the optimal risk-information cost is
\begin{align}
1-\lambda\log\max\Bigl\{\frac{1}{k_{1}}(2^{\lambda^{-1}q_{1}}-1)+1,\,\frac{1}{k_{2}}(2^{\lambda^{-1}q_{2}}-1)+1\Bigr\},\label{E:comparison}
\end{align}
and the optimal estimator is
\begin{align}
P_{\hat{Y}|X_1,X_2}^{*}(\hat{y}|x_1,x_2) & = \begin{cases}
\frac{2^{\lambda^{-1}q_1}}{2^{\lambda^{-1}q_1}+k_1-1} & \text{if}\; \hat{y}=x_1\\
\frac{1}{2^{\lambda^{-1}q_1}+k_1-1} & \text{if}\; \hat{y}\in \mathcal{Y}_1 \backslash \{x_1\}\\
0 & \text{if}\; \hat{y}\in \mathcal{Y}_2
\end{cases}
\label{E:class_optest}
\end{align}
if $\frac{1}{k_{1}}(2^{\lambda^{-1}q_{1}}-1)+1 \ge \frac{1}{k_{2}}(2^{\lambda^{-1}q_{2}}-1)+1$, and similar for the other case.
Assume $q_{1}>q_{2}$, then the optimal MAP estimate is $\bar{Y}=X_{1}$.  An estimate-compress approach would either communicate a compressed version of $\bar{Y}=X_{1}$ as in \eqref{E:class_optest}, or output any element in $\mathcal{Y}_2$ (giving a risk $1-q_{2}k_{2}^{-1}$). The risk-information cost achieved by this approach is
\begin{equation}
\min\Bigl\{ 1-\lambda\log\Bigl(\frac{1}{k_{1}}(2^{\lambda^{-1}q_{1}}-1)+1\Bigr) ,\, 1-q_{2}k_{2}^{-1}\Bigr\} 
 = 1-\lambda\log\max\Bigl\{\frac{1}{k_{1}}(2^{\lambda^{-1}q_{1}}-1)+1,\,2^{\lambda^{-1}q_{2}k_{2}^{-1}}\Bigr\}.\label{E:ec}
\end{equation}
Now, if $k_1\gg k_2$, the optimal rate constrained descriptor communicates a lossy version of $X_2$ instead, and the risk of estimate--compress in~\eqref{E:ec} is larger than \eqref{E:comparison}. 


Moreover, the gap between the rates needed by the two approaches for
a fixed risk can be unbounded. Take $q_{1}=1-q_{2}=2/3$, $k_{2}=2$,
$k_{1}\ge15$.  The minimum rate needed to achieve a risk $2/3$
is $1$ (by $\hat{Y}=X_{2}$). For the estimate-compress approach,
since $\hat{Y}\sim\mathrm{Unif}(\mathcal{Y}_{2})$ gives a risk $5/6$,
we have to compressing $X_{1}$ (by passing it through a symmetric channel with $P\{\hat{Y}=X_1\} = 1/2$) to achieve a risk $2/3$, which requires
an unbounded rate
\[
I(X;\hat{Y}) = H(\hat{Y}) - H(\hat{Y}|X_1) = \log k_{1}-\frac{1}{2}\log(k_{1}-1)-\frac{1}{2}.
\]

\begin{figure}[h!]
	\centering
	\includegraphics[width=0.65\textwidth]{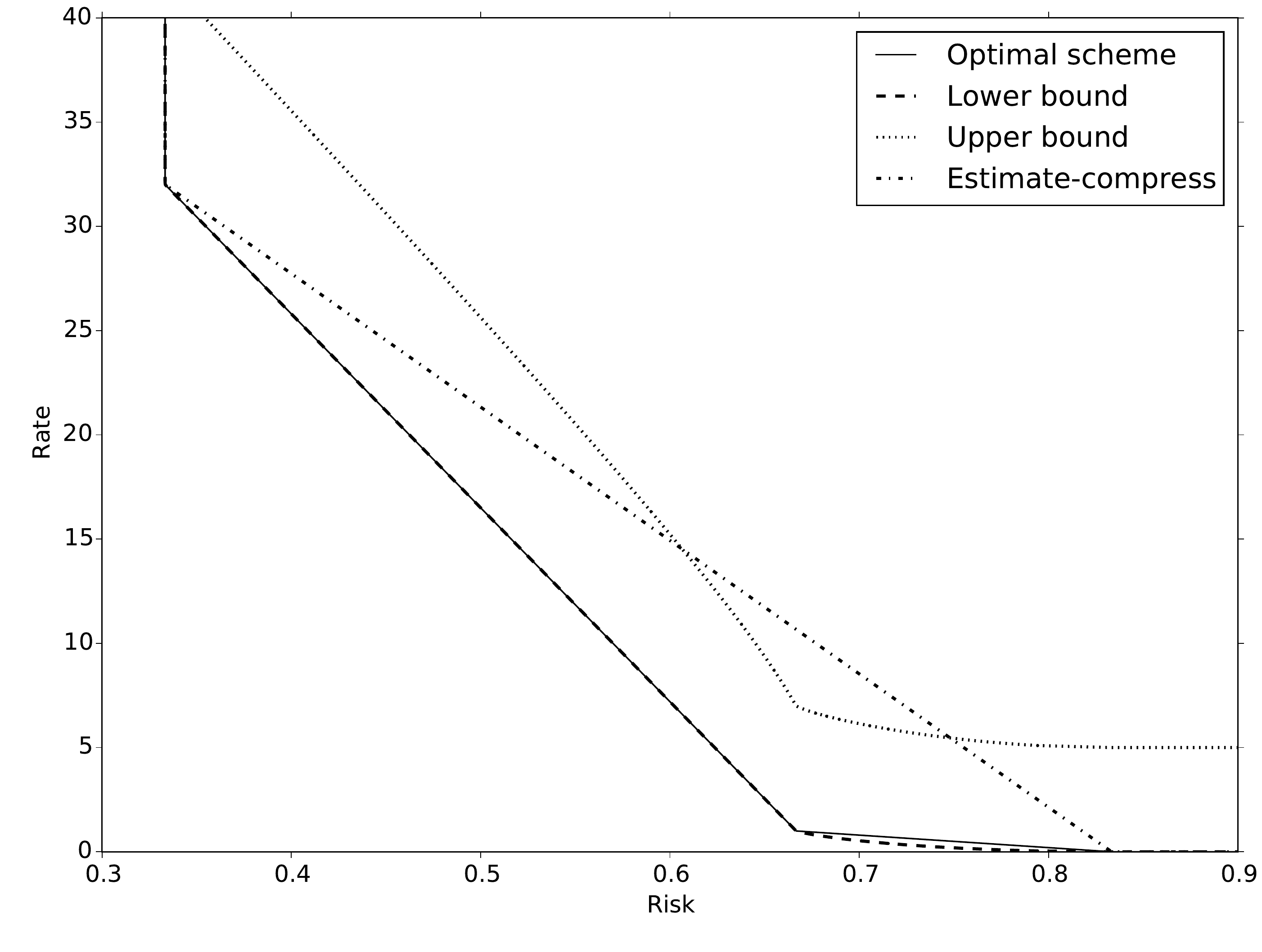}
	\caption{Tradeoff between the rate and risk in rate-constrained minimax linear classification for the optimal scheme, lower bound \eqref{E:comparison}, upper bounnd by Theorem~\ref{thm:fixedp}, and estimate-compress approach \eqref{E:ec}.}
	\label{fig:class_opt_ec}
\end{figure}

Figure~\ref{fig:class_opt_ec} compares the optimal scheme, the lower bound obtained from the optimal risk-information tradeoff~\eqref{E:comparison}, the upper bound of the optimal rate by Theorem~\ref{thm:fixedp}, and the risk-information tradeoff for the estimate-compress approach~\eqref{E:ec} for $q_{1}=1-q_{2}=2/3$, $k_{1}=2^{32}$, $k_{2}=2$. Note that the optimal scheme is to perform time sharing (using common randomness) between encoding $X_1$ using $32$ bits with risk $1/3$, encoding $X_2$ using $1$ bit with risk $2/3$, and fixing the output at one value of $X_2$ using $0$ bit with risk $5/6$. The mutual information needed by the estimate-compress approach (which is a lower bound on the actual rate needed by this approach) is strictly greater than the optimal rate (except when the risk is at its minimum $1/3$ or maximum $5/6$).

\section{Acknowledgements}
This work was partially supported by a gift from Huawei Technologies and by the  Center for Science of Information (CSoI), an NSF Science and Technology Center, under grant agreement CCF-0939370.

\bibliographystyle{IEEEtran}

\bibliography{ref}


\end{document}